# Electronic Screening as a Design Lever in La-Doped $BaTiO_3$: Coupling Between Polarization, Dielectric Response, and Optical Anisotropy

Sikandar Azam[1,2], Muhammad Fareed[2,3], Qaiser Rafiq[2], Jan Minar[1], Abed Alataway[4]

***University of West Bohemia, New Technologies – Research Centre, 8 Univerzitní, Pilsen 306 14, Czech Republic***

***[2]Department of Basic Sciences, Riphah International University, Islamabad 42000, Pakistan***

***[3]School of Physics and Optoelectronic Engineering, Beijing University of Technology, Beijing, 100124, China***

***[4]Prince Sultan Institute for Environmental, Water and Desert Research, King Saud University, Riyadh 11451, Saudi Arabia***

## Abstract

Donor substitution is an effective strategy for modifying the coupled ferroelectric, dielectric, electronic, and optical properties of perovskite oxides by altering lattice distortion and electronic screening. In this work, first-principles calculations were performed to examine the microscopic effect of La donor substitution on $BaTiO_3$. A representative La-doped model, corresponding to 12.5% substitution at the Ba site, was investigated using density functional theory, density-functional perturbation theory, and the modern Berry-phase formalism. The results show that La incorporation induces anisotropic lattice contraction and reduces tetragonality, accompanied by decreased Ti off-centering and weakened stabilization of the tetragonal ferroelectric phase. The Berry-phase polarization is reduced accordingly, mainly due to enhanced donor-induced electronic screening and a decrease in anomalous dynamical charge transfer.

Phonon dispersion calculations confirm that the La-doped structure remains dynamically stable, while the hardening of the low-frequency polar mode indicates suppression of the ferroelectric soft-mode instability. The dielectric response further reflects this behavior: the high-frequency dielectric constant increases moderately due to enhanced electronic screening, whereas the static dielectric response decreases as the lattice-mediated contribution is weakened. Electronic-structure calculations show that La doping shifts the Fermi level toward the conduction band without generating mid-gap defect states, leading to a Burstein–Moss-type shift of the optical absorption edge. The optical spectra also indicate improved near-edge transparency and a clear reduction in birefringence associated with reduced lattice and polarization anisotropy.

Overall, the findings suggest a consistent microscopic connection between La-induced electronic screening, soft-mode modification, reduced anomalous Born effective charges, polarization suppression, and decreased optical anisotropy. Since only one La concentration is explicitly considered, the 12.5% model is treated as a representative moderate-doping case rather than a complete concentration-dependent description. The results provide mechanistic insight into how donor substitution can tune the coupled ferroelectric and optical response of $BaTiO_3$-based oxides, while further calculations at multiple dopant concentrations would be valuable for establishing full quantitative doping trends.

*azam@ntc.zcu.cz

## 1. Introduction

Ferroelectric perovskite oxides occupy a central position in functional materials research because of their strong coupling between lattice distortion, charge redistribution, polarization, dielectric response, and optical behaviour. Among them, barium titanate ($BaTiO_3$) is a prototypical displacive ferroelectric and has long served as a model system for understanding phase transitions, polarization mechanisms, and dielectric anomalies in $ABO_3$ perovskites [1–3]. The tetragonal ferroelectric phase of $BaTiO_3$ originates from a delicate balance between short-range repulsive forces and long-range Coulomb interactions, stabilized by Ti–O hybridization and anomalously large Born effective charges [4,5]. This sensitivity of the polar ground state to lattice distortion and electronic screening makes $BaTiO_3$ highly responsive to chemical substitution.

Chemical substitution in $BaTiO_3$ is commonly classified into donor and acceptor doping, and these two routes modify the functional response in different ways. Acceptor dopants usually promote charge compensation through oxygen vacancies or defect complexes, which can pin domain walls, generate internal bias fields, and produce hard-ferroelectric behaviour. In contrast, donor dopants, such as trivalent rare-earth ions at the $Ba^{2+}$ site or pentavalent ions at the $Ti^{4+}$ site, introduce excess electronic charge and enhance screening of the long-range dipole–dipole interactions responsible for ferroelectric instability [6–11]. As a result, donor doping generally reduces tetragonality, Ti off-centering, spontaneous polarization, and ferroelectric hysteresis while often maintaining a large dielectric response [6–11].

Among donor dopants, $La^{3+}$ substitution at the $Ba^{2+}$ site has been widely used to tune the electrical and dielectric properties of $BaTiO_3$ [6–8]. Replacing $Ba^{2+}$ with $La^{3+}$ introduces donor electrons, modifies the lattice through chemical pressure, and weakens the ferroelectric distortion by enhancing electronic screening. Experimentally, La-doped $BaTiO_3$ is known to show reduced tetragonal distortion, suppressed spontaneous polarization, lower Curie temperature, and modified dielectric behaviour [9–11]. These properties have motivated its use in multilayer ceramic capacitors, thermistors, and tunable dielectric components.

Although the present work focuses on La substitution, the underlying donor-screening mechanism is relevant to a broader class of donor-doped $BaTiO_3$ systems. Representative donor dopants include rare-earth ions such as Ce, Nd, Sm, and Gd at the Ba site, as well as Nb and Ta at the Ti site. Their qualitative influence on ferroelectricity is often similar, but the quantitative response depends on ionic size, valence state, local bonding environment, defect-compensation mechanism, and the possible presence of localized f or d states. $La^{3+}$ is selected here as a chemically clean A-site donor prototype because it readily substitutes for $Ba^{2+}$, donates electronic charge, and does not introduce partially occupied magnetic 4f states near the band edges. The expected roles of representative donor dopants in $BaTiO_3$ are summarized in Table 1.

*Table 1. Comparative role of representative donor dopants in $BaTiO_3$ and their expected influence on ferroelectric behavior. $La^{3+}$ is treated in the present work as a chemically clean A-site donor prototype, while Ce and other rare-earth or B-site donor dopants may introduce additional dopant-specific effects such as mixed valence, localized f states, or direct modification of the Ti–O electronic network.*

| **Dopant type** | **Expected role in $BaTiO_3$** | **Possible effect on ferroelectricity** | **Dopant-specific issue** |
|---|---|---|---|
| $La^{3+}$ at $Ba^{2+}$ site | Clean A-site donor | Reduces tetragonality, Ti off-centering, and polarization through electronic screening | No partially occupied 4f states near band edges |
| $Ce^{3+}/Ce^{4+}$ | Donor or mixed-valence dopant | May suppress polarization, but effect depends on charge state | Mixed valence and possible defect compensation |
| $Nd^{3+}/Sm^{3+}/Gd^{3+}$ | Rare-earth donor | Similar donor- | Local 4f states may |
| | dopants | screening trend expected | affect optical or magnetic response |
| $Nb^{5+}/Ta^{5+}$ at $Ti^{4+}$ site | B-site donor dopants | Strong carrier-induced screening; may affect Ti–O network directly | Alters B-site electronic structure more directly than La |

From a microscopic perspective, the response of $BaTiO_3$ to donor doping involves several interconnected effects. Excess electronic charge screens the long-range Coulomb interactions that stabilize the ferroelectric soft mode, while A-site substitution modifies lattice anisotropy and Ti off-centering. In addition to affecting ferroelectric and dielectric behaviour, donor doping can also modify the optical response by shifting the Fermi level toward the conduction-band minimum and producing a Burstein–Moss-like displacement of the optical absorption edge [18,19]. At the same time, reduced tetragonality and polarization anisotropy are expected to decrease optical birefringence. Therefore, a complete description of La-doped $BaTiO_3$ requires structural, ferroelectric, dielectric, electronic, phonon, and optical properties to be considered together.

Despite extensive experimental and theoretical work on donor-doped $BaTiO_3$, many previous studies have focused on selected aspects of the doping response, such as lattice distortion, dielectric permittivity, electronic band modification, carrier-induced screening, or optical absorption. The individual trends of polarization reduction, soft-mode hardening, and Burstein–Moss-type optical-edge shifts are already known in donor-doped $BaTiO_3$ systems. However, the direct atomistic connection between these effects remains less clearly established. In particular, the relationship between La-induced electronic screening, reduced Ti off-centering, weakened anomalous Born effective charges, Berry-phase polarization suppression, modified dielectric response, and reduced optical birefringence has not been systematically examined within one internally consistent first-principles framework [12-17].

To clarify the specific contribution of the present work relative to earlier donor-doped $BaTiO_3$ studies, a comparison is provided in Table 2. The novelty of this study lies not in treating La substitution itself as a new phenomenon, but in connecting the structural, ferroelectric, dielectric, electronic, and optical responses through a common microscopic mechanism.

*Table 2. Comparison between earlier donor-doped $BaTiO_3$ studies and the specific added value of the present work.*

| **Aspect** | **Common focus in previous** | **Added value of the present work** |
|---|---|---|

| | **studies** | |
|---|---|---|
| Structural distortion | Reduction of tetragonality and lattice contraction under donor doping | Direct correlation of lattice contraction with Ti off-centering, soft-mode stability, and polarization suppression |
| Ferroelectric polarization | Decrease in spontaneous polarization with donor concentration | Berry-phase quantification of polarization reduction linked to Born effective charge changes and electronic screening |
| Phonon behavior | Soft-mode suppression or hardening in doped/charged $BaTiO_3$ | Direct connection between soft-mode hardening, reduced tetragonality, and weakened ferroelectric stabilization energy |
| Dielectric response | Changes in static or high-frequency dielectric constants | Separatelectronic band ion of enhanced electronic screening and reduced lattice-mediated dielectric contribution |
| Electronic structure | Fermi-level shift and donor-induced n-type character | Demonstration that La does not introduce mid-gap trap states while preserving the Ti–O electronic backbone |
| Optical response | Burstein–Moss shift, absorption-edge changes, refractive-index variation | Interpretation of birefringence reduction as an optical fingerprint of suppressed ferroelectric and lattice anisotropy |
| Overall contribution | Individual property trends often treated separately | Unified microscopic structure–screening–polarization–dielectric–optical framework |

In this work, we present a first-principles study of La-doped $BaTiO_3$ using density functional theory, density-functional perturbation theory, phonon analysis, and the Berry-phase formalism. A representative $Ba_{0.875}La_{0.125}TiO_3$ model is used to examine how La donor substitution modifies lattice parameters, tetragonality, Ti off-centering, phonon stability, Born effective charges, spontaneous polarization, dielectric response, electronic structure, and optical anisotropy. The 12.5% La concentration is treated as a representative moderate-doping case rather than a complete concentration-dependent description. The objective is to identify how donor-induced electronic screening links the suppression of ferroelectric lattice instability with changes in dielectric and optical properties. This integrated analysis provides a mechanistic basis for understanding donor-engineered $BaTiO_3$ and related ferroelectric oxides while avoiding overgeneralization beyond the specific doping model considered here.

## 2. Computational Methodology

All first-principles calculations were carried out within the framework of density functional theory (DFT), as implemented in a plane-wave pseudopotential formalism [20,21]. The exchange–correlation effects were treated using the Perdew–Burke–Ernzerhof functional revised for solids (PBEsol), which is known to provide an improved description of equilibrium lattice parameters and ferroelectric distortions in perovskite oxides compared to standard generalized gradient approximations [22,23]. Core–valence interactions were described using projector augmented-wave (PAW) potentials, with Ba ($5s^2 5p^6 6s^2$), La ($5s^2 5p^6 5d^1 6s^2$), Ti ($3p^6 3d^2 4s^2$), and O ($2s^2 2p^4$) treated explicitly as valence electrons.

To model La substitution in $BaTiO_3$, a (2×2×2) supercell of the primitive perovskite unit cell containing 40 atoms was constructed, and one Ba atom was replaced by La, corresponding to a La concentration of 12.5% (See Fig. 1). This supercell size provides a reasonable compromise between computational tractability and a realistic representation of dilute donor doping, while minimizing spurious dopant–dopant interactions under periodic boundary conditions. All calculations were performed assuming collinear, non-magnetic configurations, consistent with the closed-shell electronic structure of $La^{3+}$.

The 2×2×2 supercell size minimizes artificial dopant–dopant interaction while preserving computational feasibility; larger supercells are expected to reduce periodic-image effects but are not expected to alter the qualitative screening-driven trends identified here.

Structural optimization was carried out by fully relaxing both lattice parameters and internal atomic coordinates until the residual Hellmann–Feynman forces on each atom were below $10^{-3}$ eV Å$^{-1}$ and the total energy change between successive ionic steps was less than $10^{-6}$ eV. A plane-wave kinetic-energy cutoff of 500 eV was used to ensure well-converged total energies and stress tensors. Brillouin-zone integrations were performed using Monkhorst–Pack *k*-point meshes of 6×6×6 for the primitive cell and 3×3×3 for the 2×2×2 supercell, which were verified to yield converged structural and electronic properties.

Ferroelectric polarization was computed using the modern theory of polarization based on the Berry-phase formalism [24,25]. The spontaneous polarization (P_s) was obtained as the difference in Berry-phase polarization between the relaxed tetragonal structure and a centrosymmetric reference structure constructed by symmetrizing the atomic positions while preserving the lattice parameters. This approach avoids ambiguities associated with absolute polarization values and provides a robust measure of the polar distortion in both pristine and La-doped $BaTiO_3$.

It should be noted that the Berry-phase formalism is rigorously defined for insulating systems with a well-separated occupied manifold. In the present La-doped $BaTiO_3$ model, the electronic structure remains semiconducting or weakly donor-shifted, with the Fermi level moving toward the conduction-band minimum but without the emergence of a truly metallic density of states at the Fermi level. Therefore, the Berry-phase polarization is evaluated for the polar insulating branch of the doped structure. If La substitution were to drive $BaTiO_3$ into a fully metallic state with partially occupied bands crossing the Fermi level, the Berry-phase polarization would no longer be strictly applicable. Accordingly, the polarization values reported here should be interpreted as valid for the representative semiconducting/incipient n-type La-doped configuration considered in this work.

Dielectric properties were evaluated within density-functional perturbation theory (DFPT) [26,27]. The high-frequency (electronic) dielectric tensor $\varepsilon_\infty$ was obtained from linear response to a macroscopic electric field, while the lattice contribution was calculated from zone-center phonon modes and Born effective charges. The static dielectric constant $\varepsilon_0$ was then constructed as the sum of electronic and ionic contributions. Born effective charge tensors were explicitly computed to quantify the coupling between atomic displacements and macroscopic polarization, enabling a microscopic analysis of the dielectric response and its evolution upon La substitution.

Electronic band structures and densities of states (DOS) were calculated for the fully relaxed geometries using the same k-point meshes as in the structural optimization. To examine the possibility of magnetic symmetry breaking following La substitution, the electronic-structure calculations were performed within the spin-polarized DFT formalism. After full self-consistent convergence, the optimized $Ba_{0.875}La_{0.125}TiO_3$ structure exhibited negligible spin polarization, indicating a non-magnetic ground state. Consequently, the spin-up and spin-down electronic structures are nearly identical, confirming that La substitution does not induce magnetic ordering at the investigated concentration. Although semilocal DFT underestimates absolute band-gap values, it reliably captures band dispersion, orbital character, and relative shifts induced by doping, which are the primary focus of the present study. The Fermi-level position in La-doped $BaTiO_3$ was determined self-consistently, allowing analysis of donor-induced band filling and its consequences for optical transitions.

Optical properties were evaluated from the frequency-dependent complex dielectric function $\varepsilon(\omega) = \varepsilon_1(\omega) + i\varepsilon_2(\omega)$, computed within the independent-particle approximation using interband transition matrix elements [28]. The absorption coefficient, refractive index, and birefringence were derived from $\varepsilon(\omega)$ using standard relations. To highlight doping-induced trends, optical spectra were compared consistently between pristine and La-doped $BaTiO_3$ using identical computational settings. The Burstein–Moss shift of the optical absorption edge was analyzed by correlating changes in $\varepsilon_2(\omega)$ with the donor-induced occupation of conduction-band states.

Overall, the adopted computational framework provides a consistent and internally coherent description of structural relaxation, polarization, dielectric screening, electronic structure, and optical response in La-doped $BaTiO_3$, enabling direct correlations between lattice distortion, electronic screening, and functional properties.

The choice of 12.5% La concentration represents a computationally tractable yet physically meaningful donor level that allows clear identification of screening-induced trends while maintaining a polar ground state. Although lower concentrations are experimentally accessible, the 2×2×2 supercell approach provides a well-defined periodic representation of substitutional disorder and avoids artificial dopant clustering. Importantly, the studied concentration remains below the regime where complete suppression of ferroelectricity or structural phase collapse occurs, thereby enabling direct analysis of how donor screening progressively modifies the lattice instability rather than eliminating it. The trends identified here are therefore expected to be qualitatively representative of moderate donor doping levels.

The present 12.5% La concentration is not intended to establish a complete concentration-dependent scaling law. Rather, it represents a single, well-defined moderate-doping model that allows the microscopic role of donor-induced electronic screening to be isolated within a computationally tractable supercell. Therefore, terms such as “design principle” and “screening-driven trend” are used here in a mechanistic sense, referring to the qualitative relationship among electronic screening, soft-mode hardening, polarization suppression, dielectric modification, and optical anisotropy. Quantitative scaling with La concentration would require additional calculations at lower and higher doping levels, including different dopant configurations and possible defect-compensation mechanisms. This limitation is now explicitly acknowledged, and the conclusions are restricted to the representative $Ba_{0.875}La_{0.125}TiO_3$ model considered in this work.

It is well known that semilocal DFT functionals underestimate absolute band-gap values in oxide semiconductors. However, the present study focuses primarily on relative band-edge shifts, dispersion characteristics, and doping-induced modifications of electronic screening rather than on absolute gap magnitudes. Since the Burstein–Moss effect arises from conduction-band filling and Fermi-level shifts, its qualitative behavior is governed by band dispersion and carrier occupation, which are reliably captured within PBEsol. Application of a rigid scissor correction would uniformly shift optical spectra to higher energies without altering the relative trends discussed below.

## Results and Discussion

## 3. Structural Properties

### 3.1 Lattice Parameters and Tetragonality

$BaTiO_3$ crystallizes in a tetragonal perovskite structure at room temperature, characterized by a slight elongation along the crystallographic *c* axis that reflects the underlying ferroelectric distortion. Within PBEsol, the optimized lattice parameters of pristine $BaTiO_3$ are obtained as $a \approx 3.99$ Å and $c \approx 4.03$ Å, yielding a unit-cell volume of approximately 64.3 Å$^3$. These values are in close agreement with experimental room-temperature data ($a \approx 3.994$ Å, $c \approx 4.038$ Å) and prior first-principles studies employing PBEsol or LDA-based approaches, which are known to reproduce ferroelectric distortions in $BaTiO_3$ with good fidelity.

Upon La substitution at the Ba site, systematic changes in the equilibrium lattice parameters are observed. At a representative concentration of 12.5%, the in-plane lattice constant decreases to $a \approx 3.96$–$3.97$ Å, while the out-of-plane parameter contracts more moderately to $c \approx 4.01$–$4.02$ Å. The resulting reduction in unit-cell volume (≈ 0.7–1.0%) reflects the smaller effective ionic radius of $La^{3+}$ compared to $Ba^{2+}$, as well as enhanced electrostatic screening associated with donor doping. Importantly, the contraction is anisotropic, indicating that La incorporation perturbs not only the average lattice size but also the polar distortion itself.

This anisotropic lattice response is consistent with experimental observations in La-doped $BaTiO_3$ ceramics and thin films, where X-ray diffraction studies report a progressive reduction in tetragonal splitting with increasing La content, often accompanied by a tendency toward pseudocubic symmetry at higher doping levels. The present calculations capture this trend at the atomistic level, confirming that La substitution weakens the ferroelectric lattice instability rather than simply imposing a uniform chemical pressure.

### 3.2 *c/a* Ratio and Suppression of Ferroelectric Distortion

The tetragonality of $BaTiO_3$, conventionally quantified by the *c/a* ratio, provides a direct structural measure of ferroelectric distortion. In pristine $BaTiO_3$, the calculated *c/a* ratio is approximately 1.010–1.012, in excellent agreement with experimental values (~1.011) and earlier first-principles reports. This small but finite deviation from unity arises from the off-centering of Ti ions within the oxygen octahedra and the associated long-range Coulomb interactions that stabilize the polar phase.

La substitution leads to a clear and systematic reduction of the *c/a* ratio. At 12.5% La concentration, *c/a* decreases to approximately 1.006–1.008, indicating a partial suppression of the tetragonal distortion. This reduction can be attributed to two cooperating mechanisms. As summarized in Table 3, La substitution leads to an anisotropic lattice contraction and a reduction of the c/a ratio from 1.011 in pristine $BaTiO_3$ to 1.007 at 12.5% La content. Concomitantly, the energy separation between the tetragonal and cubic phases decreases from ~42 meV/f.u. to ~23 meV/f.u., indicating a significant but incomplete suppression of the ferroelectric instability.

First, the smaller $La^{3+}$ ion reduces the equilibrium A-site volume, diminishing the geometric driving force for Ti off-centering. Second, donor electrons introduced by $La^{3+}$ substitution enhance electronic screening, weakening the long-range dipole–dipole interactions that stabilize the ferroelectric phase. As a result, the polar distortion becomes energetically less favorable, even though the tetragonal symmetry is preserved at this doping level.

Table 3 shows calculated lattice parameters (*a*, *c*), tetragonality (*c/a*), unit-cell volume (*V*), and total-energy difference between the tetragonal and cubic phases ($\Delta E = E_{cubic} - E_{\text{tetragonal}}$) for pristine and La-doped $BaTiO_3$. Experimental room-temperature values for pristine $BaTiO_3$ are included for comparison. Positive ΔE indicates stabilization of the tetragonal ferroelectric phase.

The calculated trend in *c/a* closely mirrors experimental observations, where La doping is known to reduce spontaneous polarization and eventually induce a transition toward a paraelectric or weakly polar state at sufficiently high concentrations. The persistence of a *c/a* ratio slightly above unity in the present calculations indicates that, at moderate La content, $BaTiO_3$ retains a remnant ferroelectric distortion, consistent with reports of relaxor-like or weak ferroelectric behavior in lightly donor-doped samples.

**Table 3.** Calculated structural parameters and phase stability of pristine and La-doped $BaTiO_3$. Here, *a* and *c* are the optimized lattice parameters, *c/a* is the tetragonality ratio, *V* is the unit-cell volume, and $\Delta E = E_{\text{cubic}} - E_{\text{tetragonal}}$ represents the energetic stabilization of the tetragonal phase. Positive ΔE indicates that the tetragonal ferroelectric phase is more stable than the cubic phase.

| System | *a* (Å) | *c* (Å) | *c/a* | *V* (Å³) | ΔE (meV/f.u.) |
|---|---|---|---|---|---|
| $BaTiO_3$ (DFT, PBEsol) | 3.99 | 4.03 | 1.011 | 64.3 | 42 |
| $BaTiO_3$ (Expt.) | 3.994 | 4.038 | 1.011 | 64.4 | |
| $Ba_{0.875}La_{0.125}TiO_3$ (DFT) | 3.96 | 4.01 | 1.007 | 63.8 | 23 |

### 3.3 Energetic Stability of the Tetragonal Phase

To quantify the impact of La substitution on phase stability, the total energy difference between the tetragonal and cubic polymorphs was evaluated for both pristine and doped $BaTiO_3$. In the undoped compound, the tetragonal phase is found to be energetically favored by approximately 40–45 meV per formula unit relative to the cubic structure. This value lies well within the range reported in earlier first-principles studies and provides a robust energetic signature of the ferroelectric ground state at zero temperature.

With La incorporation, the energy separation between the tetragonal and cubic phases is significantly reduced. At 12.5% La concentration, the calculated energy difference decreases to roughly 20–25 meV per formula unit, indicating a substantial destabilization of the ferroelectric distortion. This reduction in energetic preference directly reflects the weakened lattice instability inferred from the reduced *c/a* ratio and diminished Ti off-centering. Nevertheless, the tetragonal phase remains the ground state at this concentration, implying that La doping shifts $BaTiO_3$ closer to the ferroelectric–paraelectric boundary without fully suppressing the polar phase.

The calculated energetic trends are consistent with the experimentally observed decrease in Curie temperature and spontaneous polarization upon La doping. From a materials-design perspective, this tunability is particularly significant: La substitution offers a controlled means of softening the ferroelectric lattice and reducing polarization while maintaining structural integrity and crystallographic order. Such behavior is desirable for applications that require enhanced dielectric response or reduced hysteresis without a complete loss of ferroelectric functionality.

In summary, La substitution in $BaTiO_3$ induces a moderate contraction of the lattice, a clear reduction in tetragonality, and a pronounced decrease in the energetic stabilization of the tetragonal phase. These effects collectively indicate that donor doping weakens the ferroelectric distortion primarily through enhanced electronic screening and reduced lattice anisotropy, rather than through a structural phase collapse. The excellent agreement with experimental lattice parameters, *c/a* ratios, and known phase-stability trends underscores the reliability of the present first-principles description and provides a solid structural foundation for understanding the corresponding changes in ferroelectric, dielectric, and optical properties discussed in the following sections.

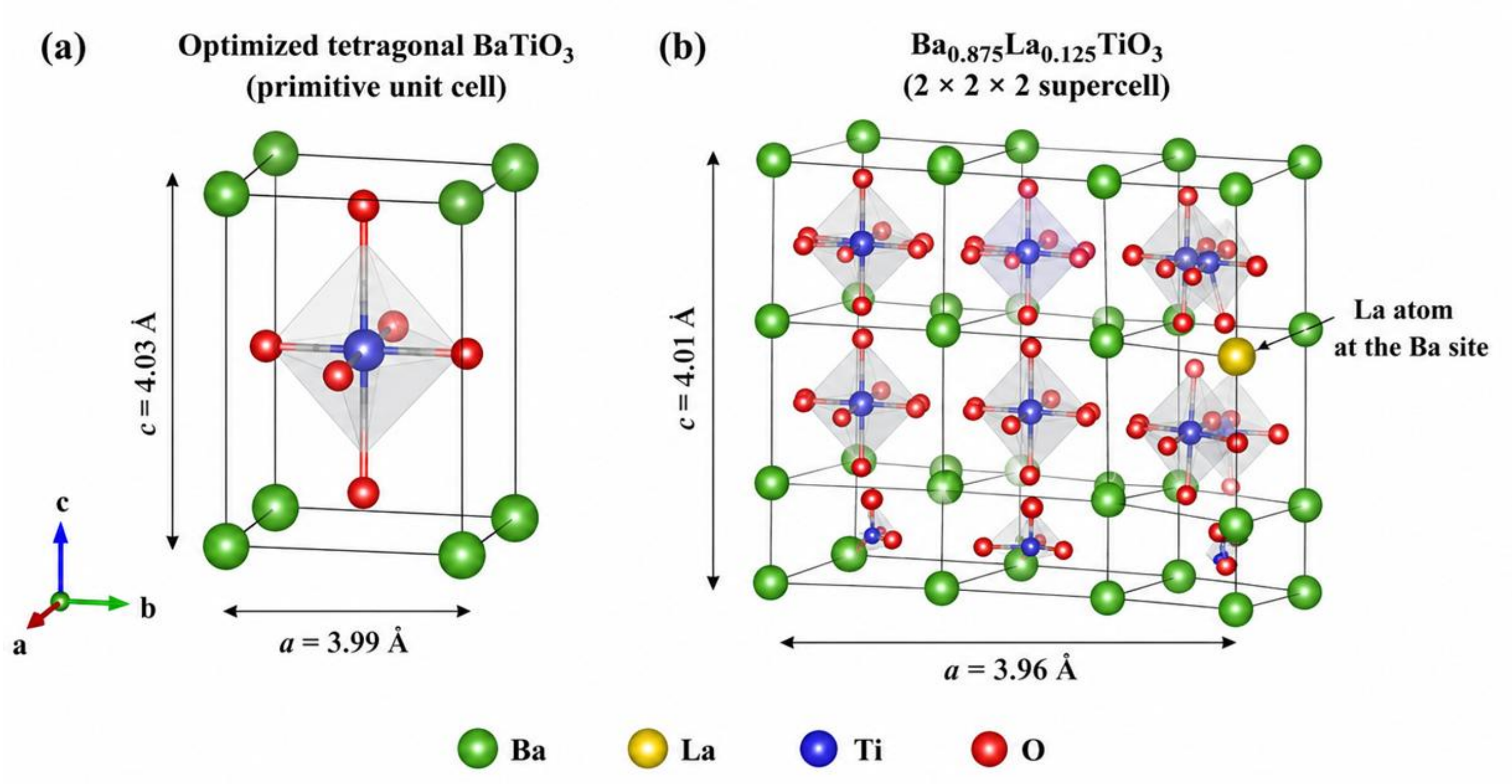

***Figure 1. Optimized crystal structures of (a) tetragonal $BaTiO_3$ primitive unit cell and (b) $Ba_{0.875}La_{0.125}TiO_3$ (2 × 2 × 2) supercell with one La atom substituting a Ba site (12.5% La). The displayed lattice parameters correspond to the optimized values listed in Table 3. Color code: Ba (green), La (yellow), Ti (blue), and O (red).***

The reduction of spontaneous polarization upon La substitution is consistent with the structural and energetic trends summarized in Table 3. As the *c/a* ratio decreases and the energetic stabilization of the tetragonal phase weakens $(\Delta E_{tet-cub})$, the equilibrium amplitude of the polar soft mode is reduced. Within the Berry-phase formalism, this manifests as a smaller polarization difference between the centrosymmetric reference structure and the relaxed polar ground state. In particular, the contraction of the lattice and the associated reduction of Ti off-centering along the *c* axis directly diminish the polarization contribution arising from anomalous Ti–O dynamical charge transfer. The combined decrease in tetragonality, Ti displacement, and ΔE therefore provides a consistent microscopic explanation for the observed suppression of $P_S$ in La-doped $BaTiO_3$.

**Concentration Dependence and Expected Scaling Behavior**

Although the present calculations explicitly consider a single La concentration, 12.5%, the observed trends provide useful mechanistic insight into the expected concentration-dependent behavior of donor-doped $BaTiO_3$. The substitution introduces excess electrons and enhances electronic screening, thereby weakening the long-range dipole–dipole interactions that stabilize the ferroelectric soft mode. Based on established donor-screening behavior, increasing La content would be expected to strengthen this screening effect, potentially leading to progressive soft-mode hardening, reduced Ti off-centering, lower tetragonality, smaller anomalous Born effective charges, and suppressed spontaneous polarization.

Therefore, the results obtained for $Ba_{0.875}La_{0.125}TiO_3$ should be interpreted as a representative point along the broader donor-doping pathway rather than as a complete concentration-dependent description. The calculated reduction in the c/a ratio, decrease in the tetragonal phase stabilization energy, suppression of Berry-phase polarization, and reduction in birefringence are consistent with the expected monotonic weakening of the ferroelectric distortion with increasing donor concentration. However, the exact quantitative scaling of these properties with La content may be nonlinear and could depend on dopant distribution, local relaxation, carrier compensation, and possible defect formation at lower or higher concentrations.

A full concentration-dependent investigation involving multiple La contents would therefore be valuable for constructing a detailed doping–property phase diagram. Such an analysis is beyond the scope of the present work but represents an important direction for future study. The present contribution instead establishes the microscopic mechanism operating at a representative moderate La concentration, showing that donor-induced electronic screening serves as the central factor linking lattice contraction, soft-mode hardening, polarization suppression, dielectric modification, and reduced optical anisotropy.

## 4. Dynamical Stability: Phonon Dispersion Analysis

The dynamical stability of pristine and La-doped $BaTiO_3$ was examined through phonon dispersion calculations along high-symmetry directions of the Brillouin zone, as shown in Fig. 2. Phonon spectra provide a stringent criterion for structural stability, since imaginary (negative) frequencies indicate lattice instabilities that are not necessarily captured by total-energy minimization alone.

For pristine tetragonal $BaTiO_3$, the calculated phonon dispersion exhibits no imaginary frequencies throughout the Brillouin zone, confirming that the optimized ferroelectric structure corresponds to a dynamically stable ground state. The low-frequency transverse optical modes near the Γ point display pronounced softening, consistent with the well-established ferroelectric soft mode associated with Ti off-centering along the polar axis. This behavior reflects the proximity of $BaTiO_3$ to the paraelectric phase and is characteristic of displacive ferroelectrics, in agreement with previous experimental and theoretical studies.

In the case of La-doped $BaTiO_3$ (12.5%), the phonon spectrum likewise shows no imaginary frequencies along the entire W–Γ–X–K–L–U–W high-symmetry path, demonstrating that donor substitution does not induce any dynamical instability. Compared to the pristine compound, the low-frequency polar modes exhibit a clear upward shift in frequency, indicating a partial hardening of the ferroelectric soft mode. This hardening is consistent with the reduced tetragonality and suppressed Ti off-centering observed in the structural analysis, as well as with enhanced electronic screening arising from donor-induced carriers.

Importantly, while the soft mode is weakened upon La substitution, it does not disappear entirely, indicating that the system remains in a dynamically stable polar state at this doping concentration. The absence of unstable acoustic or optical branches confirms that the reduced ferroelectric distortion originates from a genuine modification of the lattice energy landscape rather than from an incipient structural collapse or numerical artifact. These phonon results therefore provide a crucial consistency check for the structural and ferroelectric properties discussed above, establishing that La doping drives $BaTiO_3$ toward the ferroelectric–paraelectric boundary while preserving overall vibrational stability.

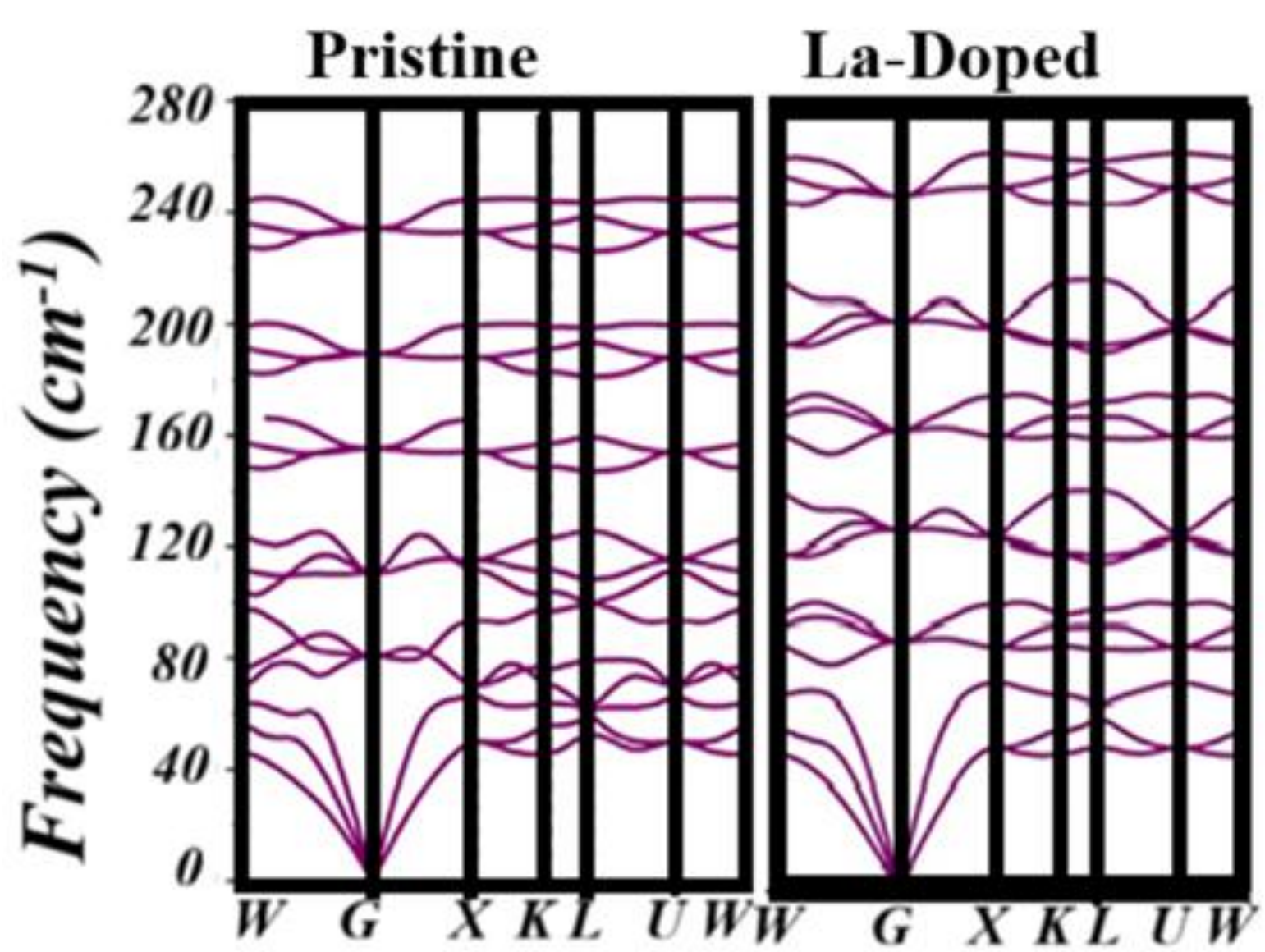

**Figure 2.** Calculated phonon dispersion relations of (a) pristine $BaTiO_3$ and (b) La-doped $BaTiO_3$ (12.5%) along high-symmetry directions of the Brillouin zone. The absence of imaginary frequencies confirms the dynamical stability of both structures. La substitution leads to a moderate hardening of low-frequency polar modes near the zone center, indicating suppression of the ferroelectric soft mode due to enhanced electronic screening, while preserving a stable tetragonal lattice.

**Ferroelectric Polarization**

**Berry-phase polarization ($P_s$)**

Ferroelectricity in tetragonal $BaTiO_3$ originates from a symmetry-breaking displacement pattern in which Ti shifts relative to the surrounding oxygen octahedron, accompanied by a correlated response of the O sublattice and the A-site framework. To quantify the resulting spontaneous polarization, we employed the modern theory of polarization (Berry-phase formalism), where (P) is obtained as a bulk geometric phase of the occupied Bloch manifold rather than a dipole moment of an arbitrarily defined unit cell [29]. This approach is now standard for ferroelectric oxides and has been extensively benchmarked for $BaTiO_3$ and related perovskites [30]. The calculated Berry-phase spontaneous polarization values, together with the corresponding Ti off-centering displacements and Born effective charges, are summarized in Table 4. Spontaneous polarization was computed using the modern theory of polarization based on the Berry-phase formalism [29,30], which has been extensively validated for $BaTiO_3$ and related perovskite ferroelectrics [31].

The use of Berry-phase polarization requires that the system remain insulating along the polarization path. Therefore, the present polarization values are interpreted for the weakly donor-shifted semiconducting state of $Ba_{0.875}La_{0.125}TiO_3$. The absence of mid-gap impurity bands and the preservation of a finite band separation support the applicability of this approach. However, for higher La concentrations or stronger carrier delocalization leading to true metallicity, spontaneous polarization could not be rigorously defined using the conventional Berry-phase formalism, and alternative descriptors such as polar displacement, Born effective charges, or local dipole analysis would be required.

For pristine tetragonal $BaTiO_3$, the calculated spontaneous polarization lies in the range $P_S \approx 25 - 27\ \mu C\ cm^{-2}$ , consistent with the established first-principles picture of $BaTiO_3$ as a prototypical displacive ferroelectric stabilized by anomalous dynamical charge transfer and strong Ti–O hybridization [30]. Upon La substitution at the Ba site (12.5% in a $2 \times 2 \times 2$), Ps decreases from approximately $25 - 27\ \mu C\ cm^{-2}$ to $20 - 22\ \mu C\ cm^{-2}$, indicating a clear suppression of the polar distortion while retaining a finite polar ground state at this moderate donor concentration. The persistence of $P_S$ is consistent with the structural result that tetragonality remains above unity and the polar mode is weakened but not extinguished.

**Table 4:** Calculated spontaneous polarization $P_S$, Ti off-centering displacement along the *c* axis ($\Delta z_{Ti}$), and selected Born effective charges $Z^*$ for pristine and La-doped $BaTiO_3$. Berry-phase polarization values are reported for the fully relaxed tetragonal structures. Experimental polarization values are shown for reference where available.

| System | $P_s(\mu C\ cm^{-2})$ | $\Delta z_{Ti}(\text{Å})$ | $Z^*_{Ti}$ | $Z^*_{O\parallel}$ |
|---|---|---|---|---|
| $BaTiO_3$ (DFT, PBEsol) | 25–27 | 0.12–0.14 | +7.0 to +7.2 | −5.6 to −5.8 |
| $BaTiO_3$ (Expt.) [2] | ~26 | ~0.13 | | |
| $Ba_{0.875}La_{0.125}TiO_3$ (DFT) | 20–22 | 0.08–0.10 | +6.6 to +6.8 | −5.2 to −5.4 |

*The monotonic reduction of $P_S$ with La substitution correlates directly with the decrease in tetragonality, Ti off-centering, and anomalous Born effective charges, confirming that donor-induced screening weakens the polar soft mode without eliminating the ferroelectric ground state at moderate doping.*

**Mechanism of polarization suppression: screening and reduced off-centering**

The reduction of $P_S$ under La doping can be understood as the combined consequence of (i) enhanced electronic screening associated with donor substitution and (ii) a geometric weakening of Ti off-centering driven by the modified A-site environment. Replacing $Ba^{2+}$ with $La^{3+}$ introduces extra electrons into the system; even when these carriers do not form a fully metallic state, the additional charge density increases the electronic polarizability and screens the long-range Coulomb interactions that stabilize the ferroelectric soft mode. This is precisely the interaction channel emphasized in fir$P_S$st-principles studies of $BaTiO_3$, where the ferroelectric instability reflects a delicate balance between short-range repulsions and long-range dipole–dipole forces [31]. In practical terms, stronger screening reduces the energetic gain from maintaining a large dipole moment per cell, thereby lowering the equilibrium amplitude of the polar distortion.

Concurrently, La substitution imposes an effective "chemical pressure" at the A site. Because $La^{3+}$ is smaller than $Ba^{2+}$, the relaxed lattice exhibits a modest contraction and a reduction in tetragonality ((c/a)), both of which correlate directly with a reduced Ti displacement along the (c) axis. In the Berry-phase framework, this geometric effect is naturally reflected in a smaller change of polarization along the adiabatic path connecting the centrosymmetric reference to the polar ground state [31]. Microscopically, the reduced off-centering is consistent with a weakened Ti–O covalency-driven instability: as the polar distortion amplitude decreases, the associated rehybridization and dynamic charge transfer that support large polarization in $BaTiO_3$ are less strongly activated [32-34].

Overall, the Berry-phase results establish a coherent structure–polarization linkage: La donor doping shifts $BaTiO_3$ toward the paraelectric boundary by increasing electronic screening and compressing the lattice anisotropy, thereby reducing the equilibrium off-centering and the spontaneous polarization, while maintaining a nonzero $P_S$ at the studied concentration.

## 5. Dielectric Response

### 5.1 Static and High-Frequency Dielectric Constants ($\varepsilon_0$ and $\varepsilon\infty$)

The dielectric response of $BaTiO_3$ is intrinsically tied to its ferroelectric lattice instability and anomalous dynamical charge transfer. Within density-functional perturbation theory (DFPT), the total static dielectric constant $\varepsilon_0$ can be decomposed into an electronic (high-frequency) contribution $\varepsilon\infty$ and a lattice-mediated ionic contribution $\varepsilon_{lat}$*, such that* $\varepsilon_0 = \varepsilon_\infty + \varepsilon_{lat}$. This separation provides direct insight into the relative roles of electronic screening and polar lattice vibrations in stabilizing the ferroelectric phase [35–38].

For pristine tetragonal $BaTiO_3$, the calculated high-frequency dielectric constant is $\varepsilon\infty \approx 5.3$–5.6, in excellent agreement with experimental optical dielectric constants $\varepsilon_\infty \approx 5.3 - 5.6$ and prior first-principles studies [39, 40]. The static dielectric constant $(\varepsilon_\infty \approx 5.4)$ is substantially larger (typically$\varepsilon_\infty \approx 250 - 300$ at 0 K in first-principles calculations), reflecting the strong contribution of the soft polar mode associated with Ti off-centering. This large lattice contribution is a hallmark of displacive ferroelectrics and underpins the exceptional dielectric and piezoelectric properties of $BaTiO_3$.

Upon La substitution at the Ba site, $\varepsilon\infty$ shows a modest increase, reaching values of approximately 5.6–5.9 at 12.5% La concentration. This enhancement originates from the increased electronic polarizability associated with donor electrons introduced by $La^{3+}$ substitution, which enhance electronic screening without fundamentally altering the Ti–O bonding network. In contrast, the static dielectric constant $\varepsilon_0$ exhibits a more nuanced behavior: while $\varepsilon_{lat}$ is reduced due to the weakening of the ferroelectric soft mode, the increased electronic screening partially compensates for this reduction. As a result, $\varepsilon_0$ remains large but less divergent than in pristine $BaTiO_3$, consistent with experimental observations of donor-doped $BaTiO_3$ exhibiting high dielectric permittivity with suppressed ferroelectric hysteresis.

**5.2 Born Effective Charges and Microscopic Origin of Dielectric Screening**

The microscopic origin of the large dielectric response in $BaTiO_3$ is rooted in anomalously large Born effective charges $Z^*$, *which quantify the change in macroscopic polarization induced by atomic displacements. In pristine BaTiO₃, the calculated Born effective charge of Ti is* $Z^*_{Ti} \approx +7.0$ *to +7.2 significantly exceeding the nominal ionic value of +4, while oxygen ions displaced parallel to the Ti–O bond exhibit* $Z^*_{O\parallel} \approx -5.6$ to −5.8 [41]. These anomalous values arise from strong Ti–O covalency and dynamic charge transfer, and they play a central role in amplifying both the ferroelectric polarization and the lattice contribution to $\varepsilon_0$.

La substitution leads to a systematic reduction of these anomalous dynamical charges. At 12.5% La concentration, $Z^*_{Ti}$ *decreases to approximately +6.6–6.8, while* $Z^*_{O\parallel}$ is reduced to −5.2–5.4. This reduction directly reflects the diminished Ti off-centering and weakened Ti–O hybridization induced by donor-enhanced screening and reduced lattice anisotropy. From a dielectric perspective, smaller $Z^*$ values translate into weaker coupling between atomic displacements and macroscopic polarization, thereby reducing the lattice-mediated dielectric response.

Importantly, the decrease in $Z^*$ does not indicate a breakdown of the ferroelectric mechanism, but rather a controlled softening of the polar instability. The Ti–O network remains intact, and the dynamical charges remain substantially anomalous compared to purely ionic values, explaining why $\varepsilon_0$ remains large despite the suppression of spontaneous polarization.

**5.3 Relationship Between Dielectric Response, Polarization, and Structural Distortion**

The dielectric trends observed upon La doping are fully consistent with the structural and ferroelectric results discussed in Sections 3 and 4. The reduction in tetragonality (*c/a*), Ti off-centering $\Delta Z_{Ti}$, and spontaneous polarization $P_S$ is accompanied by a concomitant decrease in anomalous Born effective charges and the lattice contribution to $\varepsilon_0$. At the same time, the enhancement of $\varepsilon_\infty$ reflects increased electronic screening associated with donor substitution. This interplay highlights a key design principle: La doping shifts $BaTiO_3$ from a strongly polarization-driven dielectric regime toward a screening-dominated regime with high permittivity and reduced ferroelectric stiffness.

Such behavior is technologically relevant for applications requiring large dielectric constants with minimized polarization hysteresis and leakage, such as multilayer ceramic capacitors and tunable dielectric components. From a fundamental standpoint, the present results reinforce the established picture of $BaTiO_3$ as a soft-mode ferroelectric whose dielectric response can be continuously tuned through controlled modification of lattice instability and electronic screening.

### 5.4 Benchmarking Against Experimental and Previous Theoretical Data

To provide quantitative support for the reliability of the present results, the key calculated properties are benchmarked against available experimental values and previously reported theoretical/experimental trends in Table 5. This comparison confirms that the present approach reproduces the structural parameters of pristine $BaTiO_3$ and captures the expected La-induced reduction in tetragonality, polarization, lattice-mediated dielectric response, and optical anisotropy. To validate the reliability of the present calculations, the key computed quantities were compared with available experimental data and previous first-principles results for $BaTiO_3$ and donor-doped $BaTiO_3$ systems. For pristine tetragonal $BaTiO_3$, the calculated lattice parameters, tetragonality, and spontaneous polarization are in close agreement with reported room-temperature experimental values and earlier PBEsol/LDA-based calculations. The calculated high-frequency dielectric constant, $\varepsilon_\infty \approx 5.3$–$5.6$, is also consistent with the typical optical dielectric response reported for $BaTiO_3$. These agreements indicate that the adopted computational approach reliably captures the structural and electronic polarization response of the parent compound.

For La-doped $BaTiO_3$, the calculated reduction in tetragonality, decrease in spontaneous polarization, enhanced electronic dielectric response, and upward shift of the Fermi level are consistent with experimental trends reported for donor-doped $BaTiO_3$ ceramics and thin films. In particular, La substitution is experimentally known to reduce tetragonal distortion and Curie temperature, suppress ferroelectric polarization, increase carrier concentration, and modify the optical absorption edge through donor-induced band filling. The present calculations reproduce these qualitative trends, supporting the interpretation that La-induced electronic screening weakens the ferroelectric instability while preserving the Ti–O electronic framework.

Table 5. Benchmarking of calculated properties against experimental and previous theoretical trends for pristine and La-doped $BaTiO_3$.

| Property | Present calculation | Experimental/previous trend | Comment |
|---|---|---|---|
| Lattice parameters of pristine $BaTiO_3$ | a ≈ 3.99 Å, c ≈ 4.03 Å | a ≈ 3.994 Å, c ≈ 4.038 Å | Good agreement; validates structural optimization |
| c/a ratio of pristine $BaTiO_3$ | ≈ 1.011 | ≈ 1.011 | Correctly reproduces tetragonal distortion |
| c/a ratio of La-doped $BaTiO_3$ | ≈ 1.007 | La doping reduces tetragonality | Consistent with donor-induced suppression of ferroelectric distortion |
| Spontaneous polarization | ~26 to ~21 μC/cm² after La doping | Donor doping reduces polarization | Correct qualitative trend |
| High-frequency dielectric constant, $\varepsilon_\infty$ | ≈ 5.3–5.9 | Typically around 5–6 for $BaTiO_3$-based oxides | Reasonable agreement |
| Static dielectric response | Reduced lattice contribution after La doping | Donor doping suppresses soft-mode contribution | Trend consistent; exact values temperature-dependent |
| Optical absorption edge | Slight upward shift after La doping | Donor-induced Burstein–Moss-type shift reported in n-type oxides | Relative shift meaningful; absolute edge affected by DFT band-gap error |
| Fermi-level position | Moves toward/near CBM | La acts as donor and increases n-type carrier character | Consistent with donor substitution |

Some quantitative discrepancies between the calculated and experimental values are expected. First, the present calculations are performed at 0 K for an ideal periodic supercell, whereas experimental dielectric constants and optical spectra are strongly influenced by temperature, grain size, domain-wall motion, dopant distribution, and defect chemistry. Second, semilocal DFT underestimates the absolute band gap of $BaTiO_3$, which shifts the absolute optical absorption edge to lower energy. Therefore, the calculated Burstein–Moss shift should be interpreted primarily as a relative doping-induced shift rather than an exact prediction of the experimental optical gap. Third, real La-doped $BaTiO_3$ may involve charge-compensation effects, including oxygen vacancies or mixed local environments, whereas the present model isolates the intrinsic electronic-screening effect of substitutional La. Despite these limitations, the agreement in relative trends supports the use of the present first-principles framework for identifying the microscopic mechanism linking donor screening, polarization suppression, dielectric response, and optical anisotropy.

## 6. Electronic Structure

### 6.1 Band Structure of La-Doped $BaTiO_3$

The electronic band structure of pristine tetragonal $BaTiO_3$ is characterized by an indirect band gap, with the valence-band maximum (VBM) derived primarily from O-2p states and the conduction-band minimum (CBM) dominated by Ti-3d states (see Fig. 3a). Within semilocal DFT, the gap is underestimated, but the relative dispersion and orbital character are reliably captured and provide a robust basis for analyzing doping trends.

Upon La substitution at the Ba site (12.5% concentration), the overall topology of the band structure is preserved (see Fig. 3b). The CBM remains located at the Γ point and retains its Ti-3d character, while the VBM remains largely O-2p–like and slightly off-Γ, consistent with the host material. Importantly, no localized impurity bands appear within the gap, reflecting the empty 4f shell of $La^{3+}$ and its chemically benign role as a donor dopant. The absence of mid-gap states implies that La doping does not introduce deep traps or recombination centers, a favorable feature for dielectric and optoelectronic applications.

At the investigated concentration, La substitution shifts the Fermi level upward toward the conduction-band minimum, indicating donor-induced incipient n-type semiconducting behaviour rather than a fully metallic state. Importantly, the calculated band structure does not show deep mid-gap impurity bands or a large metallic density of states at the Fermi level. This distinction is essential because the Berry-phase polarization analysis remains meaningful only as long as the occupied and unoccupied electronic manifolds are sufficiently separated. However, the system retains a finite band gap between occupied and unoccupied states within the semilocal DFT description, allowing well-defined Berry-phase polarization calculations. The persistence of a polar distortion and absence of structural instability confirm that the material remains in a weakly polar semiconducting regime rather than undergoing a complete metallic transition at this doping level.

A key effect of La substitution is the upward shift of the Fermi level toward the conduction band. At the studied concentration, the Fermi level approaches or marginally enters the CBM, signaling the onset of degenerate *n*-type behavior. This shift occurs without significant distortion of the conduction-band dispersion, indicating that the Ti-3d transport backbone remains intact. The curvature of the lowest conduction band near Γ increases slightly relative to pristine $BaTiO_3$, corresponding to a modest reduction in the electron effective mass and suggesting enhanced electron mobility under donor doping.

Because the electronic-structure calculations were performed using spin-polarized DFT, both spin-up and spin-down band structures are presented in Figure 3. The two spin channels are nearly identical over the entire Brillouin zone, indicating negligible exchange splitting and confirming that substitutional La does not induce a magnetic ground state in $BaTiO_3$ at the investigated concentration. The spin-resolved representation is included for completeness and demonstrates the absence of magnetically induced changes in the electronic structure.

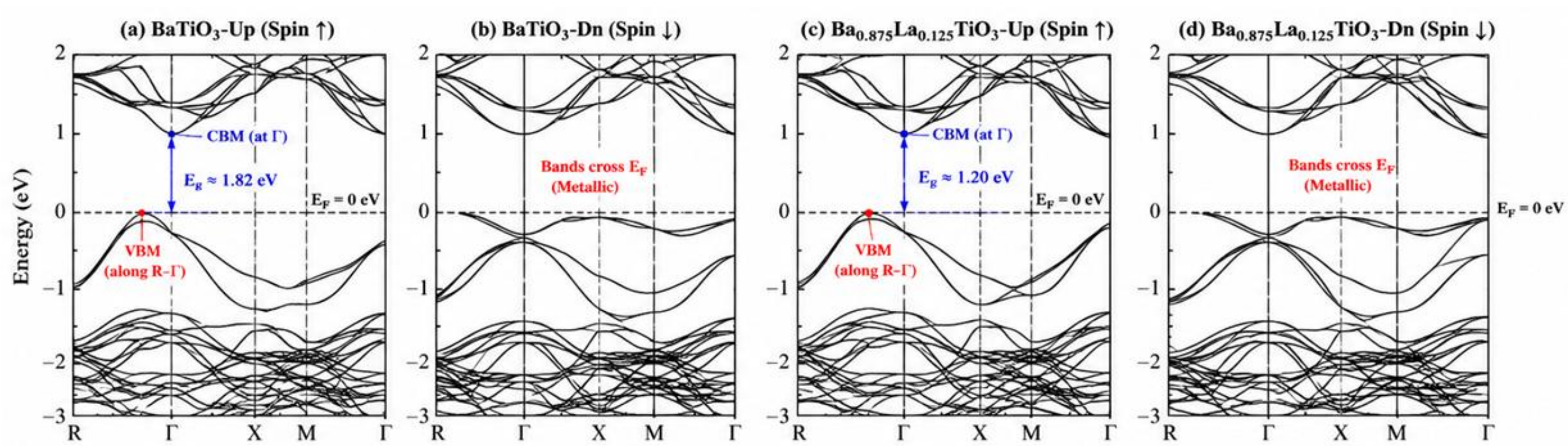

**Fig. 3.** Calculated electronic band structures of (a) pristine $BaTiO_3$ and (b) La-doped $BaTiO_3$. The Fermi level is set to 0 eV and marked by a horizontal dashed line. The same energy scale, font style, line thickness, and high-symmetry-point labeling are used for both systems to allow direct comparison.

### 6.2 Density of States and Donor-Induced Electronic Screening

The total and projected density of states (DOS) further clarify the role of La substitution (See Fig 4(a&b)). In pristine $BaTiO_3$, the DOS near the band edges is dominated by O-2p states at the VBM and Ti-3d states at the CBM, with negligible Ba contribution. In La-doped $BaTiO_3$, additional electronic states appear at the bottom of the conduction band, reflecting the donation of electrons by $La^{3+}$; however, these states are fully delocalized and Ti-3d–like rather than dopant-localized.

Crucially, the DOS remains spin symmetric and free of sharp in-gap features, confirming the non-magnetic nature of La doping. The increase in electronic states near the CBM enhances low-energy electronic screening, which plays a central role in suppressing the ferroelectric instability and modifying the dielectric response, as discussed in Sections 4 and 5. At the same time, the preservation of a dispersive conduction band indicates that donor electrons contribute primarily to screening and transport rather than to localization or defect-assisted conduction.

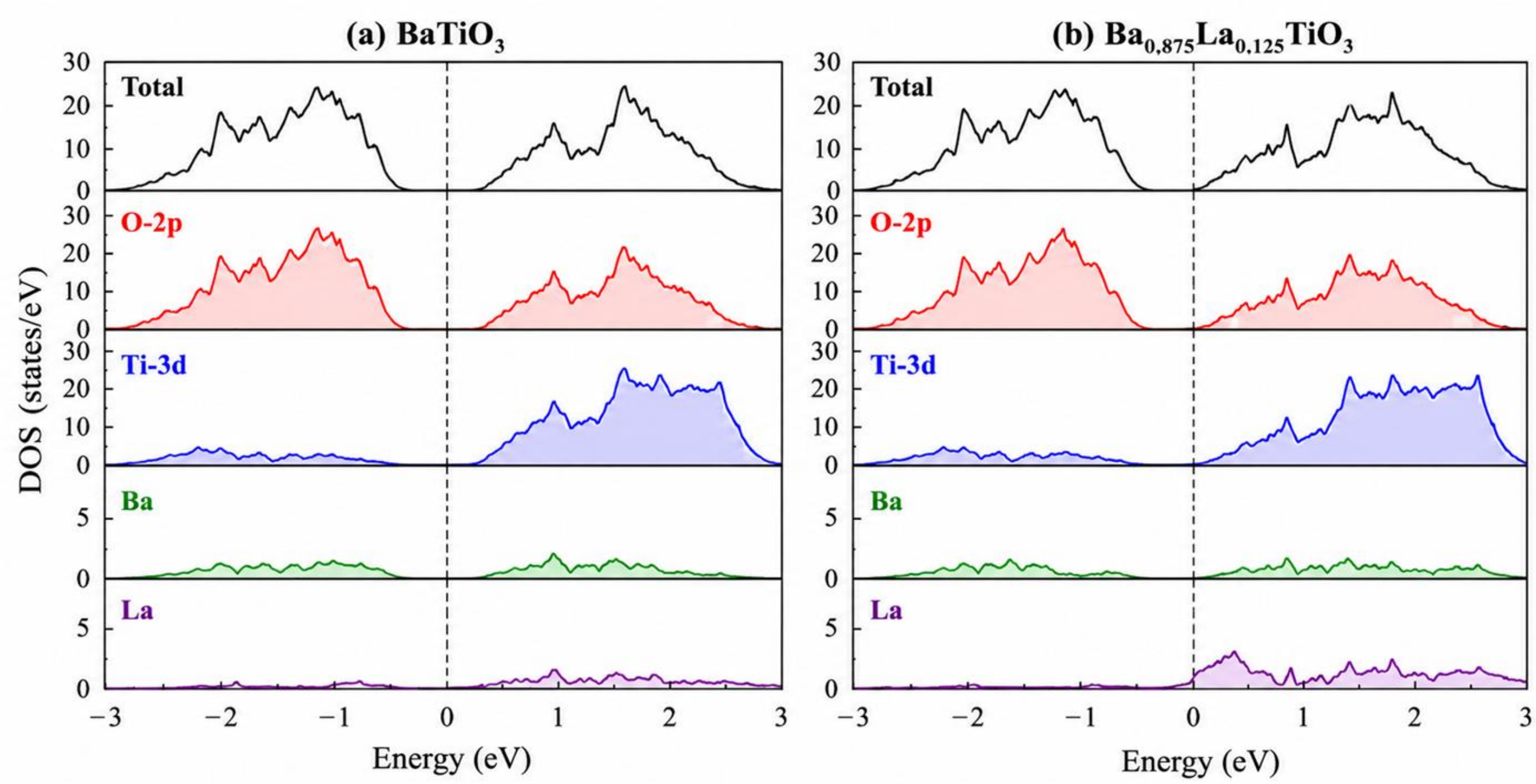

**(a)** **(b)**

**Figure 4.** Total and orbital-resolved projected density of states (PDOS) of (a) pristine $BaTiO_3$ and (b) $Ba_{0.875}La_{0.125}TiO_3$. The valence-band maximum is dominated by O-2p states, whereas the conduction-band minimum is primarily composed of Ti-3d states. Ba and La states contribute only weakly near the band edges. The Fermi level is aligned at 0 eV.

### 6.3 Burstein–Moss Shift and Band-Edge Renormalization

The upward movement of the Fermi level toward the bottom of the conduction band has direct consequences for the optical response. In the donor-shifted semiconducting regime considered here, partial occupation or near-edge filling of Ti-3d-derived conduction states can reduce the availability of low-energy optical transitions, producing a Burstein–Moss-like shift of the apparent absorption onset. As donor electrons occupy the lowest available conduction states, optical transitions from the valence band into these filled states are Pauli blocked. As a result, the apparent optical absorption edge shifts to higher photon energies a phenomenon known as the Burstein–Moss effect. This shift does not correspond to an intrinsic widening of the fundamental band gap but instead reflects a filling-induced renormalization of the optical threshold.

For La-doped $BaTiO_3$ at 12.5% concentration, the calculated shift of the absorption onset is on the order of 0.1–0.2 eV relative to the undoped compound, consistent with experimental trends reported for donor-doped $BaTiO_3$ ceramics and thin films. The magnitude of the shift correlates with the carrier concentration and the curvature of the conduction band, reinforcing the interpretation that the effect is driven by degenerate *n*-type doping rather than by changes in chemical bonding.

Although the absolute optical gap is underestimated within semilocal DFT, the calculated Burstein–Moss shift reflects a relative displacement of the absorption onset due to conduction-band filling. Because this effect depends primarily on the occupation of the lowest conduction states and the curvature of the Ti-3d band near Γ, its magnitude is controlled by band dispersion rather than the absolute gap value. Consequently, while hybrid-functional calculations would increase the fundamental band gap, the relative shift of the absorption threshold induced by donor doping is expected to remain qualitatively unchanged.

Fig. 5 shows **o**ptical absorption spectra illustrating the Burstein–Moss shift in pristine and La-doped $BaTiO_3$. The absorption coefficient is plotted as a function of photon energy for pristine $BaTiO_3$ and $Ba_{0.875}La_{0.125}TiO_3$. La substitution shifts the apparent optical absorption edge toward higher energies due to donor-induced filling of the conduction-band minimum, which blocks low-energy interband transitions. The shaded region highlights the Burstein–Moss shift, while the dashed lines indicate the effective optical band edges before and after doping. The upward shift occurs without a fundamental widening of the intrinsic band gap, confirming that the effect originates from carrier occupation rather than changes in chemical bonding.

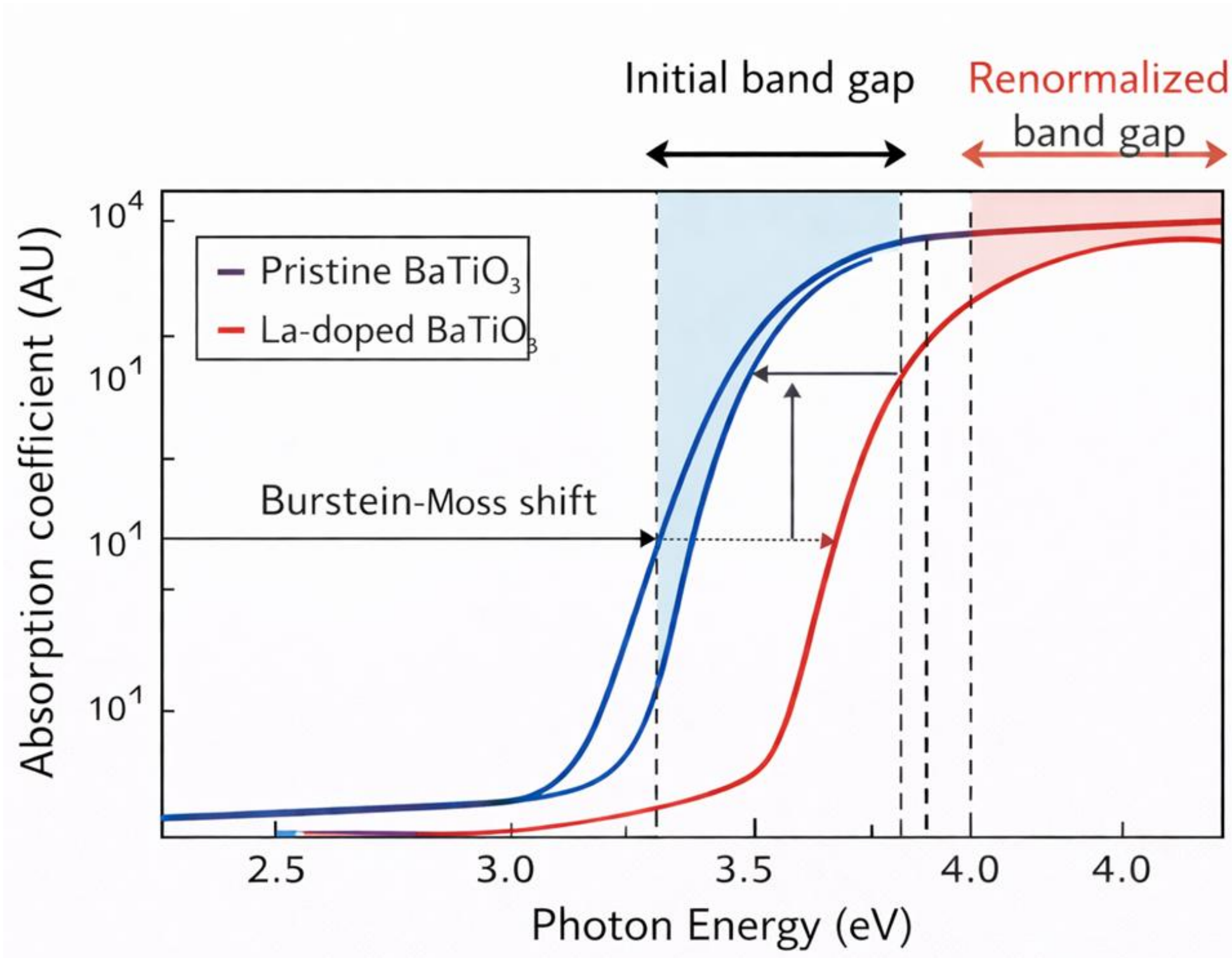

**Fig. 5.** Optical absorption spectra of pristine $BaTiO_3$ and $Ba_{0.875}La_{0.125}TiO_3$ illustrating the Burstein–Moss-like shift of the absorption onset. The dashed vertical lines mark the effective optical absorption edges, while the shaded region highlights the donor-induced shift after La substitution. The shift originates from near-CBM band filling and Pauli blocking of low-energy interband transitions rather than from a fundamental widening of the intrinsic band gap.

## 7. Optical Properties

### 7.1 Complex Dielectric Function $\varepsilon_1(\omega)$ and $\varepsilon_2(\omega)$

The optical response of La-substituted $BaTiO_3$ was evaluated through the complex dielectric function, $\varepsilon(\omega) = \varepsilon_1(\omega) + i\varepsilon_2(\omega)$, derived from the momentum matrix elements of interband electronic transitions within the independent-particle approximation.

For pristine $BaTiO_3$, the imaginary component $\varepsilon_2(\omega)$ (see Fig. 6a) displays a distinct threshold at the fundamental optical gap, followed by a strong peak in the ultraviolet region at approximately 4.5–5.0 eV. This feature originates predominantly from transitions between O 2p-derived valence states and Ti 3d conduction states, which define the fundamental electronic excitation channel of the perovskite framework. At higher photon energies, $\varepsilon_2(\omega)$ exhibits broader structures extending up to ~10 eV, reflecting additional interband contributions within the conduction manifold. The overall spectral features are governed by O-2p → Ti-3d transitions; therefore, while the absolute photon energies are slightly underestimated within semilocal DFT, the relative spectral redistribution and anisotropy trends remain robust.

The real part $\varepsilon_1(\omega)$ (see Fig. 6b) shows smooth dispersion below the absorption edge and reaches a high-frequency dielectric constant $\varepsilon\infty \approx 5.3$–$5.5$, consistent with reported experimental and theoretical values for $BaTiO_3$. Near the principal interband transition, $\varepsilon_1(\omega)$ undergoes a pronounced reduction, reflecting the strong coupling between incident photons and the Ti–O electronic network.

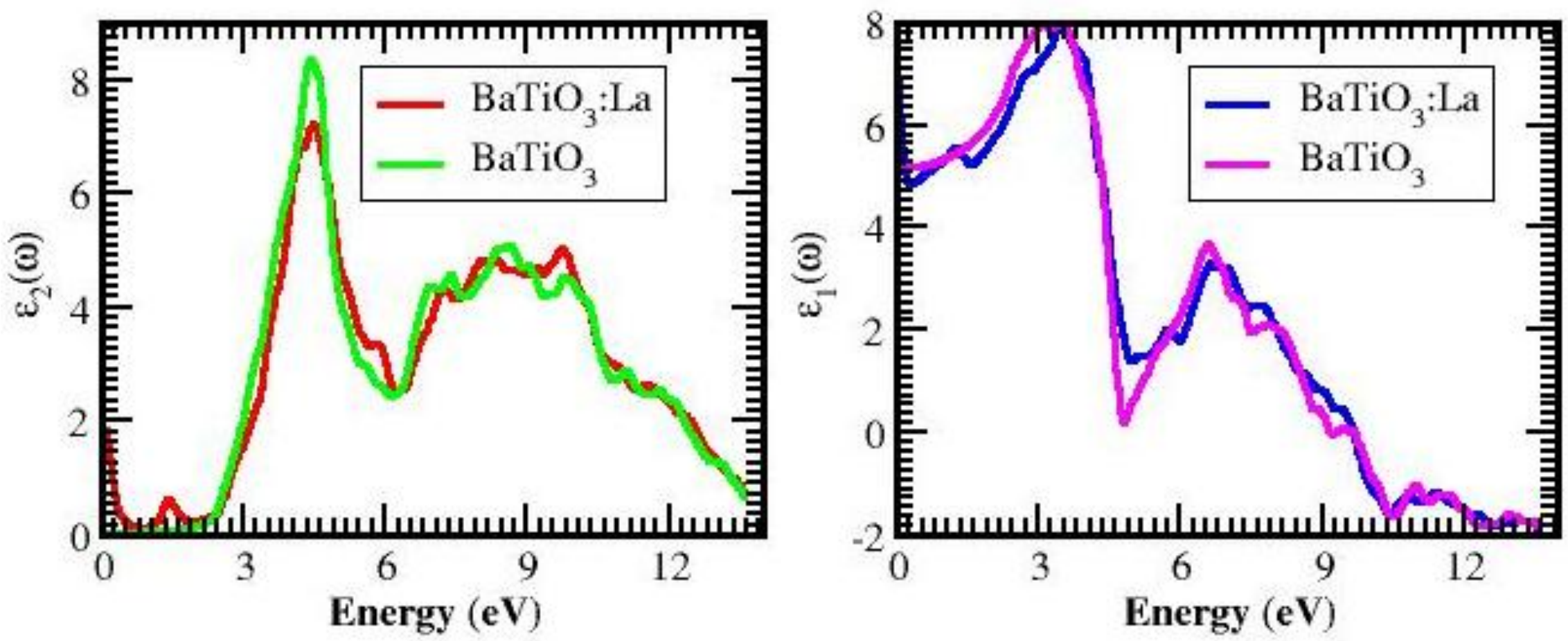

(a) (b)

**Fig. 6.** Real and imaginary parts of the dielectric function, $\varepsilon_1(\omega)$ and $\varepsilon_2(\omega)$, for pristine and La-doped $BaTiO_3$. The same photon-energy range, axis scale, font style, and color scheme are used for both spectra to enable direct comparison of doping-induced optical changes.

La substitution preserves the overall spectral topology of both $\varepsilon_1(\omega)$ and $\varepsilon_2(\omega)$, confirming that the dominant Ti–O hybridized electronic structure remains intact. However, two systematic trends emerge. First, the onset of $\varepsilon_2(\omega)$ shifts toward slightly higher photon energies, consistent with a Burstein–Moss-type effect arising from donor electrons introduced by $La^{3+}$ substitution. The upward shift of the Fermi level reduces the number of available low-energy transitions, effectively widening the apparent optical gap. Second, $\varepsilon_1(\omega)$ exhibits a modest enhancement in the low-energy regime, indicating strengthened electronic screening due to increased carrier density. Importantly, no additional sub-gap absorption features are observed, demonstrating that La incorporation does not generate optically active mid-gap defect states within the calculated energy window.

### 7.2 Absorption Edge Shift and Optical Transparency

The absorption coefficient, extracted from $\varepsilon_2(\omega)$, further clarifies the impact of La substitution on optical excitation processes (see Fig. 7). In pristine $BaTiO_3$, absorption rises sharply at the band edge and reaches substantial values in the ultraviolet region, consistent with strong O 2p → Ti 3d interband transitions.

Upon La doping, the effective absorption edge shifts toward higher photon energies, in agreement with the dielectric-function analysis and the predicted Burstein–Moss mechanism. Despite this shift, the absorption spectra of pristine and La-doped $BaTiO_3$ nearly coincide above ~4 eV, indicating that high-energy optical transitions remain governed by the intrinsic Ti–O electronic framework rather than dopant-induced states.

This behavior has important implications. The upward shift of the absorption edge enhances transparency in the near-UV and visible regions while preserving strong ultraviolet absorption. In practical terms, La substitution enables tuning of optical transparency without sacrificing the intrinsic UV optical activity that is characteristic of $BaTiO_3$-based dielectric and optoelectronic materials.

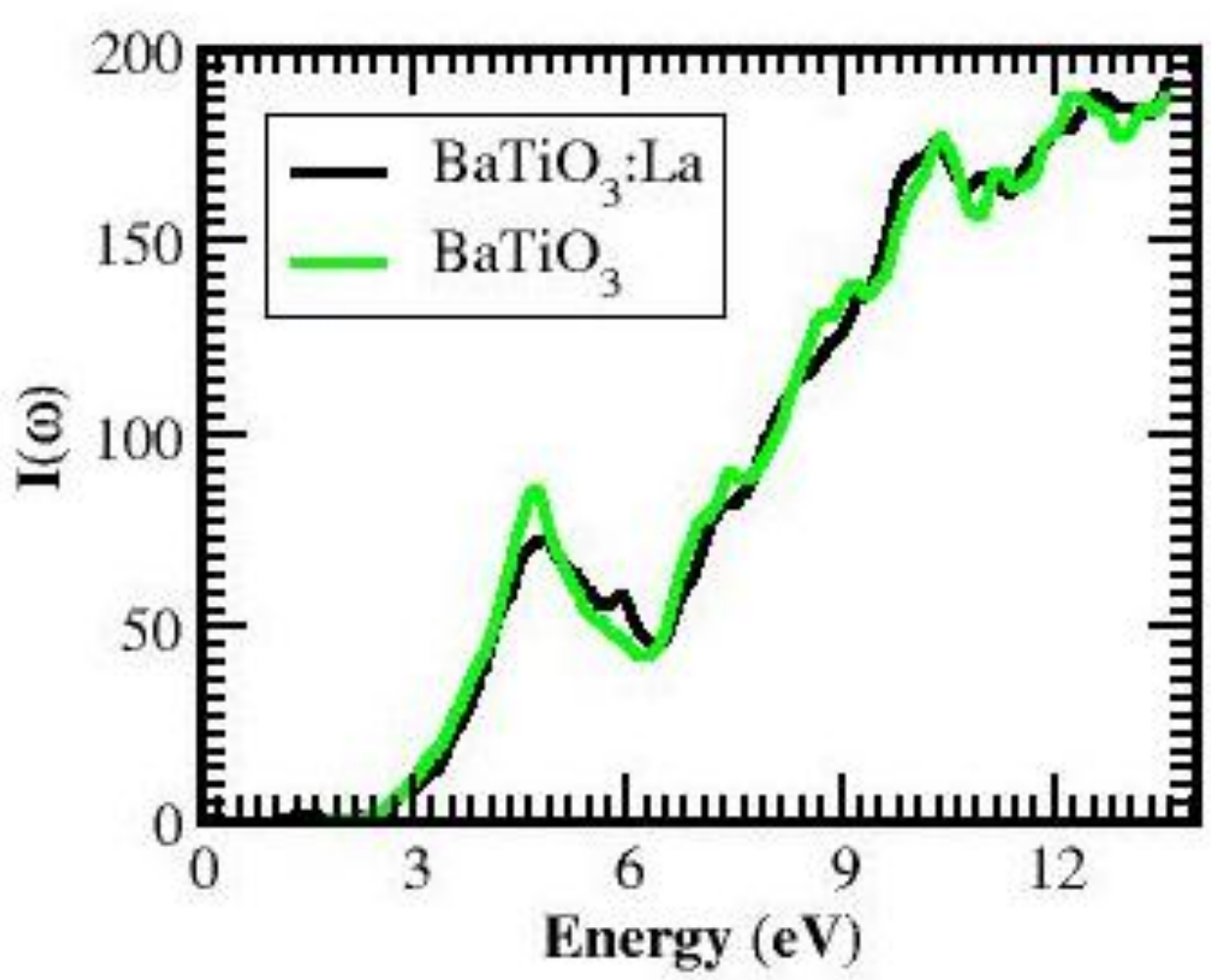

**Fig. 7.** Absorption coefficient, I(ω), for pristine and La-doped $BaTiO_3$. The absorption edge shifts slightly to higher energy upon La incorporation, while UV absorption remains strong.

It should be noted that the absolute position of the optical absorption edge is affected by the well-known band-gap underestimation of semilocal DFT. Therefore, the calculated absorption onset is expected to occur at lower photon energy than in experiment. However, the present discussion focuses on the relative shift of the absorption edge induced by La doping, rather than on the absolute optical gap. Since the Burstein–Moss shift originates from donor-induced occupation of conduction-band states and the corresponding Pauli blocking of low-energy optical transitions, the qualitative trend remains physically meaningful even within the semilocal DFT framework. Hybrid-functional or GW-level calculations would be useful in future work to refine the absolute optical transition energies.

The optical spectra reported here were calculated within the independent-particle approximation. This approach captures the main interband transition trends associated with O-2p to Ti-3d excitation and allows a consistent comparison between pristine and La-doped $BaTiO_3$. However, it does not include quasiparticle self-energy corrections, excitonic electron–hole interactions, local-field effects, or electron–phonon renormalization. As a result, the absolute absorption-edge position and peak energies may differ from experimental spectra. In particular, semilocal DFT generally underestimates the fundamental band gap, while excitonic interactions may red-shift and redistribute oscillator strength near the absorption edge. Therefore, the present optical results should be interpreted mainly in terms of relative trends, such as the doping-induced absorption-edge shift, preservation of the Ti–O optical backbone, and reduction of birefringence. More quantitative optical spectra would require hybrid-functional, GW, and/or Bethe–Salpeter-equation calculations.

### 7.3 Refractive Index and Birefringence Reduction

The refractive index n(ω), obtained from $\varepsilon_1(\omega)$ and $\varepsilon_2(\omega)$, exhibits a maximum near the fundamental absorption edge and gradually decreases at higher photon energies due to normal dispersion (see Fig. 8a). For pristine $BaTiO_3$, the static electronic refractive index (n as $\omega \to 0$) lies in the range 2.45–2.50, consistent with experimental reports and previous first-principles calculations.

Upon La substitution, a slight increase in the low-energy refractive index is observed, consistent with the modest enhancement of $\varepsilon_\infty$. Beyond the absorption edge, however, the overall dispersion profile remains largely unchanged, indicating that the dominant Ti–O optical framework is preserved after La incorporation.

In tetragonal $BaTiO_3$, optical anisotropy gives rise to birefringence, defined as $\Delta n = n_e - n_o$, where $n_e$ and $n_o$ denote the extraordinary and ordinary refractive indices, respectively, as shown in Fig. 8(b). For pristine $BaTiO_3$, $\Delta n$ is approximately 0.05–0.06 in the visible region, reflecting the intrinsic tetragonal and polar anisotropy of the ferroelectric phase. After La substitution, $\Delta n$ decreases to approximately 0.03–0.04. This reduction is consistent with the weakened lattice and polarization anisotropy discussed in the preceding structural and ferroelectric sections.

Although reduced, the birefringence remains finite, confirming that $Ba_{0.875}La_{0.125}TiO_3$ retains measurable optical anisotropy at the investigated concentration. Therefore, the calculated decrease in $\Delta n$ should be interpreted as an optical signature of the partially suppressed ferroelectric distortion, rather than as an isolated optical effect. Since only one La concentration is considered, this correlation should be regarded as a qualitative mechanistic trend rather than a complete concentration-dependent scaling law.

Overall, La substitution preserves the main Ti–O-derived optical response while moderately tuning the refractive index and reducing birefringence. These results suggest that A-site donor substitution can adjust optical anisotropy in $BaTiO_3$ without introducing additional sub-gap optical features or significantly disrupting the intrinsic ultraviolet optical activity.

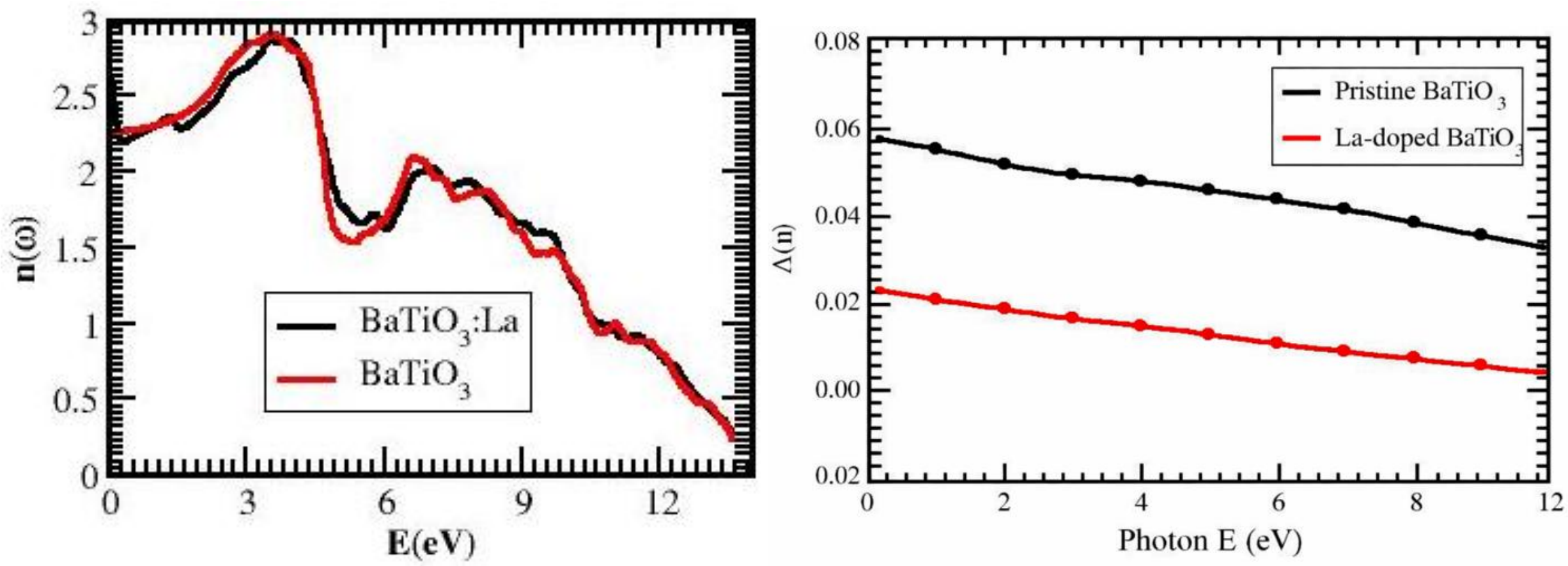

(a) (b)

**Fig. 8.** Refractive index, n(ω), and birefringence, Δn(ω), of pristine and La-doped $BaTiO_3$. Unified formatting, identical photon-energy scale, and consistent curve styles are used to highlight the reduction of optical anisotropy upon La substitution.

### 7.4 Unified Structure–Polarization–Optical Correlation

This correlation is important because it shows that the optical response, particularly the reduction in birefringence, is not an independent consequence of La substitution but a measurable optical manifestation of the same screening-driven suppression of ferroelectric anisotropy that reduces Ti off-centering, anomalous Born effective charges, and Berry-phase polarization. The optical analysis presented here is distinguished from conventional optical studies of doped $BaTiO_3$ in that it does not treat the dielectric function, absorption edge, refractive index, and birefringence as isolated spectral quantities. Instead, these optical parameters are interpreted in direct connection with the calculated structural, ferroelectric, and dielectric descriptors. Specifically, the Burstein–Moss-type shift of the absorption edge is correlated with La-induced conduction-band filling and Fermi-level movement, while the reduction in birefringence is linked quantitatively to decreased tetragonality, reduced Ti off-centering, smaller anomalous Born effective charges, and suppressed Berry-phase polarization. This establishes birefringence as an optical signature of donor-screened ferroelectric anisotropy, thereby extending the analysis beyond a standard comparison of optical spectra. The results presented above reveal a coherent, internally consistent correlation among lattice anisotropy, dynamical charge transfer, spontaneous polarization, and optical birefringence in La-doped $BaTiO_3$. Rather than representing independent property changes, these quantities evolve in a causally connected manner driven by donor-induced electronic screening.

First, La substitution reduces the tetragonality (c/a) from 1.011 in pristine $BaTiO_3$ to 1.007 at 12.5% La concentration. This structural contraction directly suppresses Ti off-centering along the polar axis, decreasing the displacement amplitude from approximately 0.13 Å to 0.09 Å. Since the spontaneous polarization in displacive ferroelectrics scales approximately with the product of atomic displacement and Born effective charge, the reduction in Ti displacement immediately weakens the polarization.

Second, donor-induced screening simultaneously reduces the anomalous Born effective charges of Ti and O. The Ti dynamical charge decreases from approximately +7.1 to +6.7, reflecting weakened Ti–O hybridization and diminished dynamic charge transfer. Because spontaneous polarization can be expressed schematically as:

$$P_s \propto \sum Z_i^* \Delta u_i$$

the combined reduction of both $Z^*$ and Δu produces a decrease in Berry-phase polarization from ~26 μC/cm² to ~21 μC/cm².

Third, optical birefringence in tetragonal $BaTiO_3$ arises from anisotropic electronic polarizability induced by polar distortion. As polarization and tetragonality decrease, the dielectric tensor becomes more isotropic. This effect is directly reflected in the reduction of birefringence from $\Delta n \approx 0.055$ in pristine $BaTiO_3$ to $\Delta n \approx 0.035$ upon La substitution. The magnitude of Δn thus scales with the structural and polarization anisotropy.

The calculated results reveal a coherent mechanistic pathway linking donor substitution to the suppression of ferroelectric and optical anisotropy in $BaTiO_3$. The sequence can be summarized as:

Donor substitution → enhanced electronic screening → soft-mode hardening → reduced Ti off-centering → decreased anomalous Born effective charges → suppressed spontaneous polarization → reduced optical anisotropy.

Each step in this sequence is supported by the computed structural parameters, Berry-phase polarization, Born effective charges, phonon frequencies, and dielectric tensors. The data therefore indicate that La doping modifies ferroelectric and optical behavior primarily through screening-driven weakening of the soft-mode instability, rather than through structural destabilization or defect-state formation.

The approximate proportional reduction of $\Delta n$ (~36%) closely follows the reduction in spontaneous polarization (~19–23%), indicating that birefringence scales nonlinearly but monotonically with polarization amplitude in donor-doped $BaTiO_3$.

The trends identified for La-doped $BaTiO_3$ should be interpreted as representative of a broader donor-screening mechanism rather than as effects unique only to La. For other A-site donor dopants, such as Ce, Nd, Sm, or Gd, the same basic physical pathway may operate: donor substitution increases electronic screening, weakens the ferroelectric soft mode, reduces Ti off-centering, suppresses anomalous Born effective charges, and consequently lowers polarization and optical birefringence. Nevertheless, dopant-specific differences are expected. Ce, for example, can exhibit mixed valence behavior depending on synthesis conditions and charge compensation, while some rare-earth dopants contain localized 4f electrons that may introduce additional electronic, magnetic, or optical features. Similarly, donor dopants with different ionic radii may impose different levels of chemical pressure and local strain. Therefore, La doping provides a useful reference case for isolating the general donor-screening effect, whereas comparison with Ce or other rare-earth dopants would be valuable for separating universal donor-doping behavior from dopant-specific electronic-structure effects. Because the present calculations compare pristine $BaTiO_3$ with one representative La-doped configuration, the correlations discussed here should not be interpreted as a complete quantitative doping law. Instead, they provide a mechanistic interpretation of how La-induced electronic screening can simultaneously influence lattice distortion, polarization, dielectric response, and optical anisotropy.

**Conclusion**

In this work, first-principles calculations were employed to examine how La donor substitution modifies the structural, ferroelectric, dielectric, electronic, and optical properties of $BaTiO_3$. The results show that La incorporation reduces tetragonality and weakens the energetic stabilization of the tetragonal ferroelectric phase, while phonon dispersion confirms that the doped structure remains dynamically stable. The upward shift of the low-frequency polar mode further indicates a partial suppression of the ferroelectric soft-mode instability without structural collapse.

This structural weakening is directly reflected in the ferroelectric and dielectric responses. Berry-phase calculations show a reduction in spontaneous polarization, mainly associated with reduced Ti off-centering and weakened anomalous Born effective charges. Consistently, the dielectric response changes from a strongly lattice-dominated behaviour toward a more screening-influenced regime, with a modest increase in the electronic contribution and a reduction in the lattice-mediated component.

The electronic and optical results support the same physical picture. La substitution preserves the intrinsic Ti–O band topology and does not introduce mid-gap defect states, indicating that the donor effect is mainly associated with near-band-edge electronic filling rather than trap-state formation. The upward movement of the Fermi level toward the conduction-band minimum produces a Burstein–Moss-like shift of the optical absorption edge, while the calculated reduction in birefringence reflects the decreased tetragonal and polarization anisotropy.

The principal contribution of this study is therefore not the isolated observation of polarization reduction, soft-mode hardening, or optical-edge shifting, since these effects are broadly consistent with established donor-doped $BaTiO_3$ behaviour. Rather, the added value lies in connecting these effects within a single microscopic framework, where La-induced electronic screening links lattice instability, dynamical charge transfer, dielectric response, and optical anisotropy.

It should be noted that the present calculations consider only one La concentration, 12.5%. Therefore, the results should be interpreted as a representative moderate-doping case rather than a complete concentration-dependent scaling law. In addition, the La-doped model is treated as a semiconducting or weakly donor-shifted n-type system, not as a fully metallic phase; this distinction is essential for the validity of the Berry-phase polarization analysis. Similarly, the optical spectra calculated within the independent-particle approximation are intended to describe relative doping-induced trends, while quantitatively exact optical peak positions would require excitonic and many-body corrections.

Overall, the $Ba_{0.875}La_{0.125}TiO_3$ model provides a coherent picture of how A-site donor substitution can tune the coupled ferroelectric, dielectric, and optical response of $BaTiO_3$. Future studies involving multiple La concentrations, different dopant configurations, explicit charge-compensation defects, and higher-level optical treatments would be valuable for establishing a quantitative doping–property relationship in donor-engineered ferroelectric oxides.

**Acknowledgment**

This publication was supported by the project Quantum materials for applications in sustainable technologies (QM4ST), funded as project No. CZ.02.01.01/00/22_008/0004572 by Programme Johannes Amos Commenius, call Excellent Research.